*Institute of Mathematics, Technical University, Wrocław*[1])

# Anisotropic Ferromagnet with Two Spins per Site

By

R. Piasecki

Green's function diagrammatic technique is used to investigate a model ferromagnet in the case, when the effective spin at a site is compounded of two anisotropically interacting electron spins. The results are: the transverse magnetic excitations ($\Delta m = \pm 1$) and the longitudinal ones ($\Delta m = 0$) are obtained in the random phase approximation, magnetization and critical temperature are calculated to the first order in $1/z$ (the reciprocal of the effective number of ions interacting with a given ion), and exchange interaction at the same site gives the contrary effects in changes of the magnitudes mentioned above in dependence on the exchange-isotropy parameter $I$ and the exchange-anisotropy parameter $D$.

С помощью диаграммной техники для температурных функций Грина рассмотрена модель ферромагнетика в случае, когда эффективный спиновый момент в узле образуется в результате сложения двух анизотропно взаимодействующих электронных спинов. Результаты: поперечные ($\Delta m = \pm 1$) и продольные ($\Delta m = 0$) магнитные возбуждения получены в приближении хаотических фаз, намагниченность и критическая температура расчитаны в приближении первого порядка по параметру $1/z$, обменное взаимодействие в узле дает противные эффекты, когда выше упомянутые величины изменяется в зависимости от параметров изотропного $I$ и анизотропного $D$ обмена.

## 1. Introduction

Systems with localized spins are described by the Heisenberg model. In most cases the effective spin $s$ is greater than 1/2. It is natural to pay attention to the effects caused by the exchange interaction of electrons on the same lattice site, particularly at temperatures much greater than 0 K and less than $T_c$, when the nearest excited states are thermally accessible. The intraatomic exchange integral for two electrons is positive [1], which leads to the Hund rule. If not only the ground state multiplet is taken into account, we shall get new effects, such as the richer spectrum of magnetic excitation (it has been mentioned in [2]), the changes of the magnetization, and those of the critical temperature. An interesting model of a ferromagnet with two spins per site and isotropic intrasite exchange integral has been considered by the authors of [3]. They used an equation-of-motion technique for double-time temperature-dependent Green's functions. The transverse magnetic excitations have been calculated with the help of various decoupling schemes. Magnetization has been obtained only in the molecular field approximation.

In this paper we present a model, which creates more possibilities of physical interpretation. The anisotropy of intrasite and intersite exchange interactions has been taken into account. Our approach has been based on the diagrammatic expansion method. The diagrammatic techniques developed by the authors [4, 5] belong to the methods applicable to more complicated spin or pseudospin systems. The first diagrammatic technique was applied to solid orthohydrogen and paradeuterium [6]; the other one was used in the spin-1 Heisenberg ferromagnet with an easy-axis single-ion anisotropy [7]. This is a limiting case for our model (when the intrasite exchange integral $I \to \infty$) and therefore, we use the same technique. This is possible because the diagrammatic technique mentioned above is simply invariant under the transition from the standard basis operators (SBO), [8] based on single-particle states to the SBO built of $n$-particle, single-ion states.

---

[1]) Wybrzeże Wyspiańskiego 27, 50-370 Wrocław, Poland.

## 2. Hamiltonian

Let us consider the model Hamiltonian of the form

$$\mathcal{H} = -\sum_{f \neq g} J_{fg}[(S^z_{f1} + S^z_{f2})(S^z_{g1} + S^z_{g2}) + \lambda(S^+_{f1} + S^+_{f2})(S^-_{g1} + S^-_{g2})] - $$
$$- \sum_f [I(\mathbf{S}_{f1} \cdot \mathbf{S}_{f2}) + 2D(S^z_{f1} S^z_{f2}) + h(S^z_{f1} + S^z_{f2})], \qquad (2.1)$$

where $S_{fi}$ are usual spin operators for $s = \frac{1}{2}$ at the $f$-th site of the $i$-th electron ($i = 1, 2$), $J_{fg} > 0$ is the exchange interaction between spins at sites $f$ and $g$, and $I > J(0)$, $D > 0$ are the isotropic and the anisotropic parts of the exchange interaction between spins at the same site, respectively, $h = g\mu_B H_z$ represents an external magnetic field applied along the $z$-direction, and $\lambda$ is the anisotropy parameter ($0 \leq \lambda \leq 1$).

We separate the model Hamiltonian into two parts,

$$\mathcal{H} = \mathcal{H}_0 + \mathcal{H}_{\text{int}}, \qquad (2.2)$$

where $\mathcal{H}_0$ is the single-ion effective-field Hamiltonian, which may be written as

$$\mathcal{H}_0 = \sum_f \mathcal{H}^f_0, \qquad (2.3)$$

and

$$\mathcal{H}^f_0 = (-2J(0)\langle S'^z \rangle - h)(S^z_{f1} + S^z_{f2}) - I(\mathbf{S}_{f1} \cdot \mathbf{S}_{f2}) - $$
$$- 2D(S^z_{f1} S^z_{f2}) - J(0)\langle S'^z \rangle^2. \qquad (2.4)$$

The interaction Hamiltonian

$$\mathcal{H}_{\text{int}} = -\sum_{f \neq g} J_{fg}[(S^z_{f1} + S^z_{f2} - \langle S'^z \rangle)(S^z_{g1} + S^z_{g2} - \langle S'^z \rangle) + $$
$$+ \lambda(S^+_{f1} + S^+_{f2})(S^-_{g1} + S^-_{g2})], \qquad (2.5)$$

where the thermal average $\langle S'^z \rangle \equiv \langle S^z_{f1} + S^z_{f2} \rangle$ will be determined self-consistently.

Noting that the operators $S^z_{f1} + S^z_{f2}$, $\mathbf{S}_{f1} \cdot \mathbf{S}_{f2}$, $S^z_{f1} \cdot S^z_{f2}$ commute with themselves, $\mathcal{H}^f_0$ is a symmetrical operator under exchange of indices, and $\mathcal{H}^f_0$ commutes with the $z$-th component of the full spin operator of a crystal, we choose as a basis the double-spin, single-ion eigenstates $|++\rangle, |+-\rangle, |-+\rangle, |--\rangle$, where $|+-\rangle \equiv |+S^z_1, -S^z_2\rangle$.

Following the formula which is in an agreement with any single-ion operator $O_f$

$$O_f = \sum_{m,n} \langle m| O_f |n\rangle L^f_{mn}, \qquad (2.6)$$

the Hamiltonian (2.4) can be expressed in terms of the SBO, such as $L^g_{mn} = |m\rangle\langle n|$, which generate transitions from the state $|n\rangle$ to the state $|m\rangle$ of the $g$-th ion.

$$\mathcal{H}^f_0 = \left(-2J(0)\langle S'^z\rangle - h - \frac{I}{4} - \frac{D}{2}\right)|++\rangle\langle++| +$$
$$+ \left(\frac{I}{4} + \frac{D}{2}\right)|+-\rangle\langle+-| - \frac{I}{2}|+-\rangle\langle-+| -$$
$$- \frac{I}{2}|-+\rangle\langle+-| + \left(\frac{I}{4} + \frac{D}{2}\right)|-+\rangle\langle-+| +$$
$$+ \left(2J(0)\langle S'^z\rangle + h - \frac{I}{4} - \frac{D}{2}\right)|--\rangle\langle--| . \qquad (2.7)$$

A common constant term $-J(0)\langle S'^z\rangle^2$ has been dropped in the energies. In order to use the Green's function diagrammatic technique of [5], we need a diagonal form of $\mathcal{H}^f_0$. After diagonalization, the Hamiltonian $\mathcal{H}^f_0$ has the following eigenstates:

$$\left.\begin{aligned}|\alpha\rangle &\equiv \frac{1}{\sqrt{2}}(|+-\rangle - |-+\rangle), & |\beta\rangle &\equiv |--\rangle, \\ |\gamma\rangle &\equiv \frac{1}{\sqrt{2}}(|+-\rangle + |-+\rangle), & |\delta\rangle &\equiv |++\rangle .\end{aligned}\right\} \qquad (2.8)$$

The eigenenergies are, respectively,

$$\varepsilon_\alpha = \frac{3I}{4} + \frac{D}{2}, \quad \varepsilon_\beta = \frac{y}{\beta} - \frac{I}{4} - \frac{D}{2}, \quad \varepsilon_\gamma = -\frac{I}{4} + \frac{D}{2}, \quad \varepsilon_\delta = -\frac{y}{\beta} - \frac{I}{4} - \frac{D}{2}, \qquad (2.9)$$

where $y = \beta(2J(0)\langle S'^z\rangle + h)$, $\beta = 1/kT$, and

$$\mathcal{H}_0 = \sum_f \sum_{i=\alpha,\beta,\gamma,\delta} \varepsilon_i L^f_{ii}. \qquad (2.10)$$

As we have in the new basis, according to (2.6),

$$\left.\begin{aligned}S^z_{j1} &= \tfrac{1}{2}(L^j_{\alpha\gamma} + L^j_{\gamma\alpha} + L^j_{\delta\delta} - L^j_{\beta\beta}), \\ S^z_{j2} &= \tfrac{1}{2}(-L^j_{\alpha\gamma} - L^j_{\gamma\alpha} + L^j_{\delta\delta} - L^j_{\beta\beta}),\end{aligned}\right\} \quad j = (f, g), \qquad (2.11)$$

and

$$\left.\begin{aligned}S^+_{f1} &= \frac{1}{\sqrt{2}}(L^f_{\alpha\beta} + L^f_{\gamma\beta} - L^f_{\delta\alpha} + L^f_{\delta\gamma}), & S^+_{f2} &= \frac{1}{\sqrt{2}}(-L^f_{\alpha\beta} + L^f_{\gamma\beta} + L^f_{\delta\alpha} + L^f_{\delta\gamma}), \\ S^-_{g1} &= \frac{1}{\sqrt{2}}(L^g_{\beta\alpha} + L^g_{\beta\gamma} - L^g_{\alpha\delta} + L^g_{\gamma\delta}), & S^-_{g2} &= \frac{1}{\sqrt{2}}(-L^g_{\beta\alpha} + L^g_{\beta\gamma} + L^g_{\alpha\delta} + L^g_{\gamma\delta}),\end{aligned}\right\} (2.12)$$

the interaction Hamiltonian (2.5) written in terms of SBO is then

$$\mathcal{H}_{\text{int}} = -\sum_{f\neq g} J_{fg}[(L^f_{\delta\delta} - L^f_{\beta\beta} - \langle S'^z\rangle)(L^g_{\delta\delta} - L^g_{\beta\beta} - \langle S'^z\rangle) + $$
$$+ 2\lambda(L^f_{\gamma\beta} + L^f_{\delta\gamma})(L^g_{\beta\gamma} + L^g_{\gamma\delta})]. \qquad (2.13)$$

Note that the SBO connected with singlet state $|\alpha\rangle$ do not appear in (2.13).

36*

### 3. Magnetic Excitations

The diagonal form of the Hamiltonian (2.10) allows to make use of the Wick-like reduction theorem [9], since in the interaction representation $L_{\mu\nu}(\tau) = e^{\mathcal{H}_0\tau} L_{\mu\nu} e^{-\mathcal{H}_0\tau}$ we get the simple $\tau$ dependence of the operators

$$L_{\mu\nu}(\tau) = e^{(\varepsilon_\mu - \varepsilon_\nu)} L_{\mu\nu} . \tag{3.1}$$

Hereafter, the diagrammatic technique developed in [5, 7] will be used to obtain the transverse ($\Delta m = \pm 1$) and the longitudinal ($\Delta m = 0$) magnetic excitations of the system. In order to find a conformity with the results for the known simpler systems, we shall assign matching priority to the "active operators" in the order: $L_{\gamma\delta}$, $L_{\beta\delta}$, $L_{\alpha\delta}$, $L_{\beta\gamma}$, $L_{\alpha\gamma}$, $L_{\beta\alpha}$; for instance

$$\langle T_\tau L^1_{\alpha\beta}(\tau_1) L^2_{\beta\alpha}(\tau_2) L^3_{\alpha\alpha}(\tau_3) \rangle_0 = \delta_{12} \mathring{G}_{\beta\alpha}(\tau_1 - \tau_2) [D_\alpha D_{\alpha\beta} + \delta_{13}(D_\alpha - D_\alpha D_{\alpha\beta})] -$$
$$- \delta_{23} \mathring{G}_{\beta\alpha}(\tau_3 - \tau_2) \delta_{13} \mathring{G}_{\beta\alpha}(\tau_1 - \tau_3) D_{\alpha\beta} , \tag{3.2}$$

where

$$\mathring{G}_{\beta\alpha}(\tau_1 - \tau_2) = e^{-(\varepsilon_\beta - \varepsilon_\alpha)(\tau_1 - \tau_2)} [\theta(\tau_1 - \tau_2) + \eta(\varepsilon_\beta - \varepsilon_\alpha)] , \quad n(\varepsilon) \equiv (e^{\beta\varepsilon} - 1)^{-1} , \tag{3.3}$$

is the noninteracting Green's function and

$$D_{\alpha\beta} \equiv D_\alpha - D_\beta , \quad D_\alpha \equiv \langle L_{\alpha\alpha} \rangle_0 \tag{3.4}$$

the thermal average. The Fourier transformation in $\tau$ is defined as follows [10]:

$$G(\tau) = \sum_n e^{-i\omega_n \tau} G(\omega_n) , \quad \omega_n = \frac{2\pi n}{\beta} ,$$

so

$$G(\omega_n) = \frac{1}{2\beta} \int_{-\beta}^{+\beta} G(\tau) e^{i\omega_n \tau} d\tau , \tag{3.5}$$

and we obtain the noninteracting Green's functions in the Fourier space. Certain vertices of Green's function lines carry the respective weight factors.

$$\left.\begin{aligned}
\mathring{G}_{\beta\alpha}(\omega_n) &= \frac{1}{\beta} \frac{1}{y/\beta - I - D - i\omega_n}, & D_{\alpha\beta} &= Z^{-1}(e^{-\beta(I+D)} - e^{-y}), \\
\mathring{G}_{\beta\gamma}(\omega_n) &= \frac{1}{\beta} \frac{1}{y/\beta - D - i\omega_n}, & D_{\gamma\beta} &= Z^{-1}(e^{-\beta D} - e^{-y}), \\
\mathring{G}_{\alpha\delta}(\omega_n) &= \frac{1}{\beta} \frac{1}{y/\beta + I + D - i\omega_n}, & D_{\delta\alpha} &= Z^{-1}(-e^{-\beta(I+D)} + e^y), \\
\mathring{G}_{\gamma\delta}(\omega_n) &= \frac{1}{\beta} \frac{1}{y/\beta + D - i\omega_n}, & D_{\delta\gamma} &= Z^{-1}(-e^{-\beta D} + e^y),
\end{aligned}\right\} \tag{3.6}$$

where $Z = e^{-\beta D}(e^{-\beta I} + 1) + 2 \cosh y$.

Let us first consider the transverse Green's function

$$K^{+-}_{f_1 g_1}(\tau) = \tfrac{1}{2} \langle T_\tau S^+_{f_1}(\tau) S^-_{g_1}(0) \rangle . \tag{3.7}$$

According to (2.12), the $q - \omega_n$ component of (3.7) is a sum of 16 Green's functions written in the language of SBO. In the chain approximation, which is equivalent to

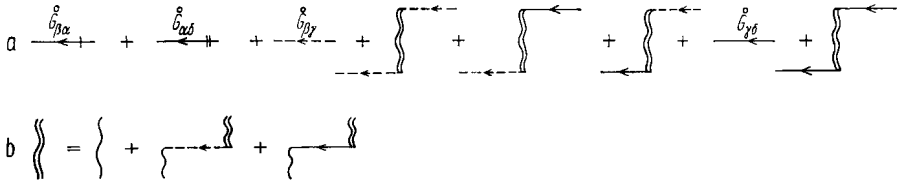

Fig. 1. Chain diagrams for the transverse Green's function and the chain renormalized transverse interaction

RPA, only six Green's functions, such as given below, give a contribution different from zero:

$$K_{11}^{+-}(q, \omega_n) = \tfrac{1}{4} (\langle T_\tau L_{\alpha\beta}^f(\tau) L_{\beta\alpha}^g(0)\rangle_{q,\omega_n} + \langle T_\tau L_{\delta\alpha}^f(\tau) L_{\alpha\delta}^g(0)\rangle_{q,\omega_n} +$$
$$+ \langle T_\tau L_{\gamma\beta}^f(\tau) L_{\beta\gamma}^g(0)\rangle_{q,\omega_n} + \langle T_\tau L_{\gamma\beta}^f(\tau) L_{\gamma\delta}^g(0)\rangle_{q,\omega_n} +$$
$$+ \langle T_\tau L_{\delta\gamma}^f(\tau) L_{\beta\gamma}^g(0)\rangle_{q,\omega_n} + \langle T_\tau L_{\delta\gamma}^f(\tau) L_{\gamma\delta}^g(0)\rangle_{q,\omega_n}). \qquad (3.8)$$

The diagrams are shown in Fig. 1a. The chain-like diagrams can be summed up with the help of the renormalized transverse interaction defined in Fig. 1b.

Hence, we obtain

$$2\beta\lambda \tilde{J}_q^t(\omega_n) = \frac{2\beta\lambda J(q) (i\omega_n - D - y/\beta)(i\omega_n + D - y/\beta)}{(i\omega_n - \omega_q^+)(i\omega_n - \omega_q^-)}, \qquad (3.9)$$

where

$$\omega_q^\pm = 2J(0)\langle S'^z\rangle + h - \lambda J(q) D_{\delta\beta} \pm$$
$$\pm \sqrt{(\lambda J(q) D_{\delta\beta})^2 + D^2 - 2\lambda J(q) D(D_{\delta\gamma} - D_{\gamma\beta})}. \qquad (3.10)$$

Thus, to this order of approximation we have

$$K_{11}^{+-}(q, \omega_n) = -\frac{1}{4\beta}\left(\frac{D_{\alpha\beta}}{i\omega_n - (y/\beta - I - D)} + \frac{D_{\delta\alpha}}{i\omega_n - (y/\beta + I + D)} +\right.$$
$$\left.+ \frac{(D_{\gamma\beta} + D_{\delta\gamma})(i\omega_n - 2J(0)\langle S'^z\rangle - h) + D(D_{\delta\gamma} - D_{\gamma\beta})}{(i\omega_n - \omega_q^+)(i\omega_n - \omega_q^-)}\right).$$
$$(3.11)$$

The analytic continuation to the whole complex $\omega$ plane by $i\omega_n \to \omega + i\delta$ gives four poles of the Green's function at $\omega_q^\pm$, $\omega_3$, $\omega_4$, which are identified as the elementary excitation energies of the system. Notice that $\langle S'^z\rangle$ should be calculated to the zeroth order in $1/z$, (the consistency in each order has been discussed in detail by the authors of [7]), and $D_{\delta\beta} = \langle S'^z\rangle$, $D_{\delta\gamma} - D_{\gamma\beta} = 3\langle (S'^z)^2\rangle_0 - 2$, therefore (3.10) can be rewritten as

$$\omega_q^\pm = h + \langle S'^z\rangle_0 [2J(0) - \lambda J(q)] \pm$$
$$\pm \sqrt{[\lambda J(q) \langle S'^z\rangle_0]^2 + D^2 - 2\lambda J(q) D[3\langle (S'^z)^2\rangle_0 - 2]}. \qquad (3.12)$$

It is interesting that within the considered approximation only $\omega_q^\pm$ describe the collective excitations of the spin system, while the latter excitations,

$$\omega_3 = 2J(0)\langle S'^z\rangle + h - I - D, \qquad \omega_4 = 2J(0)\langle S'^z\rangle + h + I + D, \quad (3.13)$$

have a local character. They are effective-field excitations between the triplet states $|\beta\rangle$, $|\delta\rangle$, and the singlet state $|\alpha\rangle$, because the interaction Hamiltonian (2.13) contains

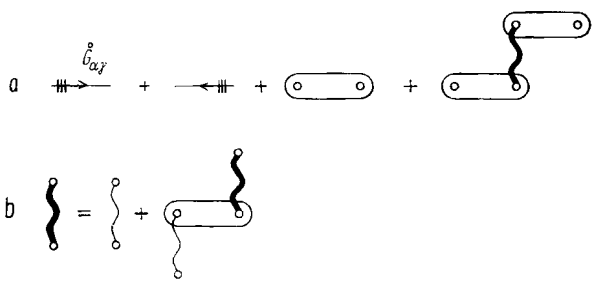

Fig. 2. Chain diagrams for the longitudinal Green's function and the chain renormalized longitudinal interaction

no matrix elements between these levels. A higher-order approximation is needed to create dispersion in the poles (3.13). These excitations could be thought of as a kind of "optical" spin wave. By setting $D = 0$, we recover the results of [3]. On the other hand, for $I \gg J(0)$, one can expect a behaviour of the spin system similar to that of spin $s' = s_1 + s_2 = 1$. In this case, the elementary excitations are described by the full Green's function

$$K_{fg}^{+-}(\tau) = \tfrac{1}{2} \langle T_\tau \big(S_{f1}^+(\tau) + S_{f2}^+(\tau)\big) \big(S_{g1}^-(0) + S_{g2}^-(0)\big)\rangle, \tag{3.14}$$

and after simple reductions, we find

$$K^{+-}(q, \omega_n) = -\frac{1}{\beta} \frac{D_{\delta\beta}\big(i\omega_n - 2J(0)\langle S'^z\rangle - h\big) + D(D_{\delta\gamma} - D_{\gamma\beta})}{(i\omega_n - \omega_q^+)(i\omega_n - \omega_q^-)}. \tag{3.15}$$

The results of [7] are given at the limit $I \to \infty$. Now, let us consider the longitudinal Green's function in the chain approximation

$$K_{f1g1}^{zz}(\tau) = \langle T_\tau \big(S_{f1}^z(\tau) - \langle S_{f1}^z\rangle\big)\big(S_{g1}^z(0) - \langle S_{g1}^z\rangle\big)\rangle. \tag{3.16}$$

Using (2.11) and taking into account that the interaction Hamiltonian (2.13) does not contain the operators connected with singlet state we find

$$K_{11}^{zz}(q, \omega_n) = \tfrac{1}{4} \langle T_\tau L_{\alpha\gamma}^f(\tau) L_{\gamma\alpha}^g(0)\rangle_{0, q, \omega_n} + \langle T_\tau L_{\gamma\alpha}^f(\tau) L_{\alpha\gamma}^g(0)\rangle_{0, q, \omega_n} + \\ + \langle T_\tau \big(L_{\delta\delta}^f(\tau) - L_{\beta\beta}^f(\tau) - \langle S'^z\rangle\big)\big(L_{\delta\delta}^g(0) - L_{\beta\beta}^g(0) - \langle S'^z\rangle\big)\rangle. \tag{3.17}$$

The diagrams are shown in Fig. 2a. After summing up with the help of longitudinal interaction defined in Fig. 2b, we get

$$K_{11}^{zz}(q, \omega_n) = \frac{1}{4\beta}\left(\frac{D_{\gamma\alpha}}{i\omega_n + I} - \frac{D_{\gamma\alpha}}{i\omega_n - I}\right) + \frac{D_\delta + D_\beta - D_{\delta\beta}^2}{1 - 2\beta J(q)(D_\delta + D_\beta - D_{\delta\beta}^2)}\frac{\delta_{n0}}{4}, \tag{3.18}$$

where

$$D_{\gamma\alpha} = Z^{-1}(1 - e^{-\beta I})e^{-\beta D}. \tag{3.19}$$

Since for excitations $\omega_n \neq 0$, only the first two terms describe the transitions between the triplet state $|\gamma\rangle$ and the singlet state $|\alpha\rangle$. In the considered approximation they have a local character, too. Taking into account the full Green's function for $I \gg J(0)$ we obtain

$$K^{zz}(q, \omega_n) = \frac{(D_\delta + D_\beta - D_{\delta\beta}^2)\delta_{n0}}{1 - 2\beta J(q)(D_\delta + D_\beta - D_{\delta\beta}^2)}. \tag{3.20}$$

The result of [11] is given at the limit $D \to 0$.

### 4. Magnetization

The magnetization $\langle S'^z \rangle$ is given by $\langle L_{\delta\delta} \rangle - \langle L_{\beta\beta} \rangle \equiv P_\delta - P_\beta$. In the zeroth order in $1/z$ $P_\delta = D_\delta$, $P_\beta = D_\beta$, and hence

$$\langle S'^z \rangle = 2 \sinh y (2 \cosh y + e^{-\beta D} \Delta)^{-1}, \tag{4.1}$$

where $\Delta \equiv e^{-\beta I} + 1$. From the condition $I > J(0)$, we have for $\Delta$ the inequality $1 < \Delta < e^{-\beta J(0)} + 1$. The result in the MFA for $s = 1$ is given at the limit $I \to \infty$.

The first-order correction to $P_\delta$ is given by the four diagrams shown in Fig. 3. (an additional constant term $I/4 - D/2$ has been added to the energies of (2.9) as only the difference in energies is essential in our calculations).

Similar diagrams have been used in [7], but in this paper after summing over the frequency in each diagram, we obtain more general expressions

$$\left.\begin{array}{l}
\text{a)} \;\; [\hat{A}] \dfrac{1}{N} \sum\limits_q \dfrac{\beta J(q)}{1 - 2\beta J(q)(D_\delta + D_\beta - D_{\delta\beta}^2)}, \\[2mm]
\text{b)} \;\; [\hat{D}] \dfrac{1}{N} \sum\limits_q 2\beta\lambda J(q) \left\{ \dfrac{1}{2} + \dfrac{\omega_q^+ - \varepsilon_\beta}{\omega_q^+ - \omega_q^-} \eta(\omega_q^+) - \dfrac{\omega_q^- - \varepsilon_\beta}{\omega_q^+ - \omega_q^-} \eta(\omega_q^-) \right\}, \\[2mm]
\text{c)} \;\; [\hat{G}] \dfrac{1}{N} \sum\limits_q 2\beta\lambda J(q) \left\{ \dfrac{1}{2} + \dfrac{\omega_q^+ + \varepsilon_\delta}{\omega_q^+ - \omega_q^-} \eta(\omega_q^+) - \dfrac{\omega_q^- + \varepsilon_\delta}{\omega_q^+ - \omega_q^-} \eta(\omega_q^-) \right\}, \\[2mm]
\text{d)} \;\; [\hat{K}] \left(-\dfrac{1}{N}\right) \sum\limits_q \dfrac{2\lambda J(q)}{\omega_q^+ - \omega_q^-} \left\{ \dfrac{\omega_q^+ - \varepsilon_\beta}{\omega_q^+ + \varepsilon_\delta} \eta(\omega_q^+) - \dfrac{\omega_q^- - \varepsilon_\beta}{\omega_q^- + \varepsilon_\delta} \eta(\omega_q^-) \right\} + \eta(-\varepsilon_\delta),
\end{array}\right\} \tag{4.2}$$

where $[\hat{A}] \equiv D_\delta(1 - D_\delta - D_\beta - 2D_{\delta\beta} + 2D_{\delta\beta}^2)$, $[\hat{D}] \equiv D_\delta(1 - D_{\delta\gamma})$, $[\hat{G}] \equiv -D_\delta D_{\gamma\beta}$, $[\hat{K}] \equiv -D_{\delta\gamma}$. The first-order correction to $P_\beta$ is given by three diagrams obtained by interchanging $L_{\delta\delta}$ with $L_{\beta\beta}$ in the diagrams of Fig. 3a, b, c, and by diagram (d'). Thus

$$\left.\begin{array}{l}
\text{a')} \;\; [\hat{B}] \dfrac{1}{N} \sum\limits_q \dfrac{\beta J(q)}{1 - 2\beta J(q)(D_\delta + D_\beta - D_{\delta\beta}^2)}, \\[2mm]
\text{b')} \;\; [\hat{E}] \dfrac{1}{N} \sum\limits_q 2\beta\lambda J(q) \left\{ \dfrac{1}{2} + \dfrac{\omega_q^+ - \varepsilon_\beta}{\omega_q^+ - \omega_q^-} \eta(\omega_q^+) - \dfrac{\omega_q^- - \varepsilon_\beta}{\omega_q^+ - \omega_q^-} \eta(\omega_q^-) \right\}, \\[2mm]
\text{c')} \;\; [\hat{H}] \dfrac{1}{N} \sum\limits_q 2\beta\lambda J(q) \left\{ \dfrac{1}{2} + \dfrac{\omega_q^+ + \varepsilon_\delta}{\omega_q^+ - \omega_q^-} \eta(\omega_q^+) - \dfrac{\omega_q^- + \varepsilon_\delta}{\omega_q^+ - \omega_q^-} \eta(\omega_q^-) \right\}, \\[2mm]
\text{d')} \;\; [\hat{L}] \left(-\dfrac{1}{N} \sum\limits_q \dfrac{2\lambda J(q)}{\omega_q^+ - \omega_q^-} \left\{ \dfrac{\omega_q^+ + \varepsilon_\delta}{\omega_q^+ - \varepsilon_\beta} \eta(\omega_q^+) - \dfrac{\omega_q^- + \varepsilon_\delta}{\omega_q^- - \varepsilon_\beta} \eta(\omega_q^-) \right\} - \eta(\varepsilon_\beta) \right),
\end{array}\right\} \tag{4.3}$$

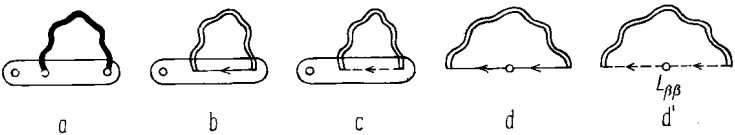

Fig. 3. First-order correction diagrams for $P_\delta$. Small circles which are not connected with the interaction lines are the $L_{\delta\delta}$ vertices; those which attach to interaction lines are the $L_{\delta\delta} - L_{\beta\beta} - \langle S'^z \rangle$ vertices

where $[\hat{B}] \equiv D_\beta(1 - D_\delta - D_\beta + 2D_{\delta\beta} + 2D_{\delta\beta}^2)$, $[\hat{E}] \equiv -D_\beta D_{\delta\gamma}$, $[\hat{H}] \equiv -D_\beta(1 + D_{\gamma\beta})$, $[\hat{L}] \equiv D_{\gamma\beta}$. Finally, we obtain the general formula for the magnetization,

$$\langle S'^z \rangle = D_{\delta\beta} + D_{\delta\beta}(1 - 3D_\delta - 3D_\beta + 2D_{\delta\beta}^2) \frac{1}{N} \sum_q \frac{\beta J(q)}{1 - 2\beta J(q)(D_\delta + D_\beta - D_{\delta\beta}^2)} +$$

$$+ [\hat{D} - \hat{E}] \frac{1}{N} \sum_q \frac{2\beta\lambda J(q)}{\omega_q^+ - \omega_q^-} \{(\omega_q^+ - \varepsilon_\beta) \eta(\omega_q^+) - (\omega_q^- - \varepsilon_\beta) \eta(\omega_q^-)\} +$$

$$+ [\hat{G} - \hat{H}] \frac{1}{N} \sum_q \frac{2\beta\lambda J(q)}{\omega_q^+ - \omega_q^-} \{(\omega_q^+ + \varepsilon_\delta) \eta(\omega_q^+) - (\omega_q^- + \varepsilon_\delta) \eta(\omega_q^-)\} +$$

$$+ [\hat{K}]\left(-\frac{1}{N}\right) \sum_q \frac{2\lambda J(q)}{\omega_q^+ - \omega_q^-} \left\{\frac{\omega_q^+ - \varepsilon_\beta}{\omega_q^+ + \varepsilon_\delta} \eta(\omega_q^+) - \frac{\omega_q^- - \varepsilon_\beta}{\omega_q^- + \varepsilon_\delta} \eta(\omega_q^-)\right\} -$$

$$- [\hat{L}]\left(-\frac{1}{N}\right) \sum_q \frac{2\lambda J(q)}{\omega_q^+ - \omega_q^-} \left\{\frac{\omega_q^+ + \varepsilon_\delta}{\omega_q^+ - \varepsilon_\beta} \eta(\omega_q^+) - \frac{\omega_q^- + \varepsilon_\delta}{\omega_q^- - \varepsilon_\beta} \eta(\omega_q^-)\right\} +$$

$$+ \eta(-\varepsilon_\delta) + \eta(\varepsilon_\beta) . \tag{4.4}$$

In the case of an isotropic ferromagnet ($\lambda = 1$, $D = 0$) with spin $s = 1$ ($I \to \infty$) the magnetization reduces to the formula of [11], (see: page 80). It is interesting that for $D = 0$, $\varepsilon_\delta = \omega_q^+ = -\varepsilon_\beta$, and we must again sum up over frequency the diagrams in Fig. 3d for $P_\delta$ and d' for $P_\beta$.

## 5. Curie Temperature

Setting the external magnetic field equal to zero and considering the small $\langle S'^z \rangle$ limit in the equation $\langle S'^z \rangle = P_\delta - P_\beta$, according to [7], we find the Curie temperature

$$\frac{1}{2\beta_c J(0)} = \frac{2}{\gamma_A} + \frac{2}{\gamma_A}\left(1 - \frac{6}{\gamma_A}\right) \frac{1}{N} \sum_q \frac{\beta_c J(q)}{1 - 4\beta_c J(q) \gamma_A^{-1}} +$$

$$+ \frac{2}{N} \sum_q \frac{e^{-\beta_c m_A}}{(1 - e^{-\beta_c m_A})^2} (1 - 2\beta_c \lambda J(q) \gamma_A^{-1})^2 +$$

$$+ \frac{1}{N} \sum_q \frac{2\beta_c \lambda J(q) [D(\gamma_A - 2)(\varDelta + 2) \varDelta^{-1} - 2\lambda J(q)]}{\gamma_{mA}^2} \frac{1 + e^{-\beta_c m_A}}{1 - e^{-\beta_c m_A}} -$$

$$- 2 e^{-\beta_c D}(e^{-\beta_c D} - 1)^{-2} , \tag{5.1}$$

where $\gamma_A = 2 + \varDelta e^{-\beta_c D}$, $m_A^2 = D^2 + (4D\lambda J(q)/\varDelta)(1 - (2 + \varDelta)/\gamma_A)$. The result in the MFA is given by keeping only the first term on the right-hand side of the equation; the other terms represent the first-order correction. For weak anisotropy, $D \ll J(0)$, to first order in $D$, formula (5.1) reduces to

$$\frac{1}{2\beta_c J(0)} = f_0(\varDelta) + f_1(\varDelta) \frac{1}{N} \sum_q \frac{\beta_c \lambda J(q)}{1 - 4\beta_c \lambda J(q)(2 + \varDelta)^{-1}} +$$

$$+ f_2(\varDelta) \frac{1}{N} \sum_q \frac{\beta_c J(q)}{1 - 4\beta_c J(q)(2 + \varDelta)^{-1}} +$$

$$+ f_3(\varDelta) \frac{1}{N} \sum_q \frac{[\beta_c \lambda J(q)]^2}{1 - 4\beta_c \lambda J(q)(2 + \varDelta)^{-1}} + f_4(\varDelta) \frac{1}{N} \sum_q [\beta_c \lambda J(q)]^2 +$$

$$+ \beta_c D \left\{ f_5(\Delta) + f_6(\Delta) \frac{1}{N} \sum_q [\beta_c \lambda J(q)]^2 + f_7(\Delta) \frac{1}{N} \sum_q \frac{\beta_c \lambda J(q)}{1 - 4\beta_c \lambda J(q)(2+\Delta)^{-1}} + \right.$$

$$+ f_8(\Delta) \frac{1}{N} \sum_q \frac{\beta_c J(q)}{1 - 4\beta_c J(q)(2+\Delta)^{-1}} + f_9(\Delta) \frac{1}{N} \sum_q \frac{[\beta_c \lambda J(q)]^2}{1 - 4\beta_c \lambda J(q)(2+\Delta)^{-1}} +$$

$$+ f_{10}(\Delta) \frac{1}{N} \sum_q \frac{[\beta_c \lambda J(q)]^2}{[1 - 4\beta_c \lambda J(q)(2+\Delta)^{-1}]^2} + f_{11}(\Delta) \frac{1}{N} \sum_q \frac{[\beta_c J(q)]^2}{[1 - 4\beta_c J(q)(2+\Delta)^{-1}]^2} +$$

$$\left. + f_{12}(\Delta) \frac{1}{N} \sum_q \frac{[\beta_c \lambda J(q)]^3}{[1 - 4\beta_c \lambda J(q)(2+\Delta)^{-1}]^2} \right\}, \qquad (5.2)$$

where the form of the functions $f_i(\Delta)$, $i = 0, \ldots, 12$ is given in the Appendix. Our formula is a generalization of the one obtained in [7] in which a calculation error has been found, i.e. in our case the sum equals $f_7 + f_8 = -20/27$, whereas in the mentioned paper it equals $-21/27$. In the other extreme limit, for $D \gg I > J(0)$, we expand the terms in (5.2) in powers of $D^{-1}$ and get

$$\frac{1}{2\beta_c J(0)} = 1 - \frac{1}{N} \sum_q \frac{2\beta_c J(q)}{1 - 2\beta_c J(q)} - \frac{1}{N} \sum_q \frac{\beta_c [\lambda J(q)]^2}{D} -$$

$$- \frac{1}{N} \sum_q \frac{\beta_c [\lambda J(q)]^3}{D^2} + O(D^{-3}, e^{-\beta_c D}). \qquad (5.3)$$

For $J(0)/D \to 0$ and $I/D \to 0$, the states $|\gamma\rangle$, $|\alpha\rangle$ are suppressed and we recover the result obtained in [12], for an Ising system of spin 1/2 (with an exchange parameter $4J_{fg}$). On the other hand, setting $\lambda = 0$ in (2.1), we obtain the Ising system with two spins $s = 1/2$ per site and the equation for the Curie temperature

$$\frac{1}{2\beta_c J(0)} = \frac{2}{\gamma_\Delta} + \frac{2}{\gamma_\Delta} \left(1 - \frac{6}{\gamma_\Delta}\right) \frac{1}{N} \sum_q \frac{\beta_c J(q)}{1 - 4\beta_c J(q) \gamma_\Delta^{-1}}. \qquad (5.4)$$

Setting $D/J(0) \to \infty$ we recover again the result for the Ising system of spin 1/2.

## 6. Discussion

The numerical computations of the collective transverse excitations for a simple cubic lattice have shown that in the case of anisotropic intrasite exchange interaction the spin-wave energy is lowered with respect to the one for the limiting case $I \to \infty$.

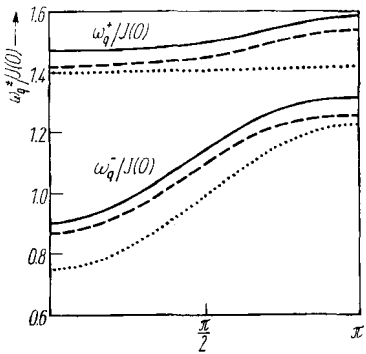

Fig. 4. Collective spectra for a simple cubic lattice along the (001) direction calculated from formula (3.12) for $kT/J(0) = 1.0$, $\lambda = 0.5$. The solid lines represent the $\omega_q^\pm/J(0)$ for spin-1 Heisenberg ferromagnet ($I \to \infty$) with easy-axis single-ion anisotropy for $D/J(0) = 0.2$. The $\omega_q^\pm/J(0)$ for our model for $I/J(0) = 2.0$ are plotted for $D/J(0) = 0.2$ (dashed lines) and for $D/J(0) = 0.05$ (dotted lines). ——— $\langle S'^z \rangle_0 = 0.791$, – – – 0.762, ⋯⋯ 0.719

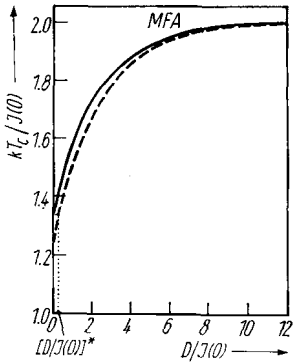 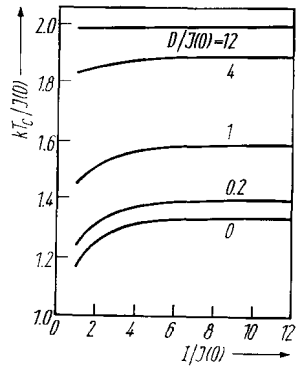

Fig. 5　　　　　　　　　　　　　　　　　Fig. 6

Fig. 5. Plot of Curie temperature versus anisotropy strength for spin-1 Ising ferromagnet with easy-axis single-ion anisotropy (solid line) and for our model for $I/J(0) = 2.0$ (dashed line)

Fig. 6. Plot of $kT_c/J(0)$ versus $I/J(0)$ for various values of $D/J(0)$

In the other case, when $I =$ const and $D$ increases, the spin-wave energy increases, too (see Fig. 4). The general formula (4.4) takes into account the fluctuations of spontaneous magnetization. If we consider the Curie temperature, then in the MFA the existence of the effects mentioned below can already be seen (see Fig. 5). Let us denote by $T_c^1$ the critical temperature of our model, by $T_c^2$ that of the anisotropic Heisenberg ferromagnet with spin $s = 1$, by $T_c^3$ that of the isotropic Heisenberg ferromagnet $s = 1$, by $T_c^4$ that of the isotropic model with two spins per site. The double inequality takes place: $T_c^2 > T_c^3 > T_c^4$. If $I$ is determinate ($I = I_1$), then there exists such a value of $[D/J(0)]^*$ that for $[D/J(0)]^* < [D/J(0)]$ there is $T_c^2 > T_c^1 > T_c^3$; for $[D/J(0)]^* > [D/J(0)]$ there is $T_c^3 > T_c^1 > T_c^4$, $T_c^1 \to T_c^2$ in the limiting case $I \to \infty$. Furthermore, Fig. 6 confirms the qualitative conclusion that for small values of $I/J(0)$, $T_c^1$ is mostly decreasing for low values of $D/J(0)$. It is interesting that after simple modifications, e.g. $f$, $g$ denote sites in plane 1 and plane 2, respectively, $I < 0$, ..., the model Hamiltonian (2.1) can describe the system of two ferromagnetic layers coupled by weak antiferromagnetic interaction. This approximation has been used in [13] to describe CoCl$_3$. The two-layer-type magnets have been considered in [14], too. Calculations are underway.

*Acknowledgements*

The author is grateful to Dr. K. Walasek for his continuous interest and encouragement. I would like to thank Dr. H. Konwent, who referred me to the papers [1, 3], and to L. Budzianowski (M.Sc.) for numerical computations.

**Appendix**

$$f_0 = \frac{2}{2 + \Delta}, \quad f_1 = \frac{-4(8 - \Delta + 2\Delta^2)}{3(2 + \Delta)^3}, \quad f_2 = \frac{-2(8 + 2\Delta - \Delta^2)}{(2 + \Delta)^3},$$

$$f_3 = \frac{16\Delta(1 - \Delta)}{(2 + \Delta)^4}, \quad f_4 = \frac{-4}{3(2 + \Delta)^2}, \quad f_5 = \frac{2\Delta}{(2 + \Delta)^2},$$

$$f_6 = \frac{-8\varDelta}{3(2+\varDelta)^3}, \qquad f_7 = \frac{2(20 - 4\varDelta - 13\varDelta^2 - 12\varDelta^3)}{3(2+\varDelta)^4},$$

$$f_8 = \frac{-2\varDelta(20 + 8\varDelta - \varDelta^2)}{(2+\varDelta)^4}, \qquad f_9 = \frac{-32\varDelta(2 - 3\varDelta + \varDelta^2)}{3(2+\varDelta)^5},$$

$$f_{10} = \frac{8(16 - 10\varDelta + 5\varDelta^2 - 2\varDelta^3)}{3(2+\varDelta)^5}, \qquad f_{11} = \frac{-8\varDelta^2(10 - \varDelta)}{(2+\varDelta)^5},$$

$$f_{12} = \frac{-32\varDelta(2 - 3\varDelta + \varDelta^2)}{(2+\varDelta)^6}.$$